\documentclass[11pt]{article}
\usepackage{amsmath,amsfonts,amssymb,amsthm}
\usepackage[margin=1in]{geometry}
\usepackage[colorlinks]{hyperref}
\usepackage{graphicx}

\newcommand{\ket}[1]{|{#1}\rangle}

\begin{document}


\title{Exploring quantum properties of bipartite mixed states under coherent and incoherent basis }

\author{Sovik Roy$^{1}$, Anushree Bhattacharjee$^{1}$
\thanks{s.roy2.tmsl@ticollege.org} 
\thanks{a.bhattacharya.tmsl@ticollege.org}\\
$^1$  Department of Mathematics, Techno Main Salt Lake (Engg. Colg.), \\Techno India Group, EM 4/1, Sector V, Salt Lake, Kolkata  700091, India\\\\
Chandrashekhar Radhakrishnan$^{2}$
\thanks{chandrashekar@physics.iitm.ac.in}\\
$^2$  Centre for Quantum Information, Communication and Computing (CQuICC),\\
Indian Institute of Technology Madras, Chennai 600036, India\\\\
Md. Manirul Ali$^{3}$
\thanks{manirul@citchennai.net}\\
$^3$  Centre for Quantum Science and Technology, \\Chennai Institute of Technology, Chennai 600069, India\\\\
Biplab Ghosh$^{4}$ 
\thanks{biplab1976@gmail.com}\\
$^4$ Department of Physics, \\Vivekananda College for Women, Barisha, Kolkata - 700008, India\\\\
}

\maketitle

\begin{abstract}
\noindent 
Quantum coherence and quantum entanglement are two different manifestations of the superposition principle.  In this article we show that the right choice of 
basis to be used to estimate coherence is the separable basis.  The quantum coherence estimated using the Bell basis does not represent the coherence in the 
system, since there is a coherence in the sytem due to the choice of the basis states.  We first compute the entanglement and quantum coherence in the two qubit 
mixed states prepared using the Bell states and one of the states from the computational basis. The quantum coherence is estimated using the $\ell_{1}$-norm of 
coherence, the entanglement is measured using the concurrence and the mixedness is measured using the linear entropy.  Then we estimate these quantities in 
the Bell basis and establish that coherence should be measured only in separable basis, whereas entanglement and mixedness can be measured in any basis.  
We then calculate the teleportation fidelity of these mixed states and find the regions where the states have a fidelity greater than the classical teleportation fidelity.  
We also examine the violation of the Bell-CHSH inequality to verify the quantum nonlocal correlations in the system.  The estimation of the above mentioned quantum correlations, 
teleportation fidelity and the verification of Bell-CHSH inequality is also done for bipartite states obtained from the tripartite systems by the tracing out of one of 
their qubits.  We find that for some of these states teleportation is possible even when the Bell-CHSH inequality is not violated, signifying that nonlocality is 
not a necessary condition for quantum teleportation. 

Pacs No.: 03.65.Ud ~~\and \and ~~03.67.Mn \and \and ~~03.67.-a
\end{abstract}


\section{Introduction:}\label{sec:introduction}
In quantum mechanics, superposition gives rise to two interesting phenomena, one of which is coherence and the other is known as 
entanglement\cite{glauber1963,sudarshan1963}. Although the entanglement and coherence are two different manifestations of the superposition principle, 
they are different in their nature and properties.  The quantum entanglement measures the non-local correlations in the systems and the quantum coherence 
is a measure of the total quantumness in the system. While quantum entanglement, when measured in any reference basis, retains its trait, quantum coherence is a basis dependent quantity
\cite{wootters1998,baumgratz2014}. This has generated an upsurge of interest on coherence and, for the last several years the exploration of this feature 
has gained momentum and has been studied in many different areas, alongside the study of entanglement
\cite{bromley2015,chitambar2016,radhakrishnan2016,radhakrishnan2019,radhakrishnan2020}. A considerable amount of research works on 
entanglement of bipartite and as well as of multipartite systems have already been done\cite{thew2001, thew2001a, munro2001, roy2010}. 
The focus of this article will be on bringing coherence and entanglement together in a bipartite two qubit mixed state scenario and to observe how these two features together contribute to the capacity of transferring information via such mixed states in a protocol such as teleportation. Such a capacity of states is, called teleportation fidelity.

Another interesting fact in the class of mixed states is that there exists a mixed entangled state, known as Werner state, which does not violate Bell's inequality\cite{werner1989}. 
Moreover Werner state can be used as a quantum teleportation channel (average optimal teleportation fidelity exceeding $\frac{2}{3}$) 
even without violating the Bell-CHSH inequality\cite{roy2010}.  We know Bell inequality marks the boundary between the classical and quantum natures. 
The relation between Bell violation and entanglement has been studied already for a class of two qubit maximally entangled mixed states 
\cite{thew2001,thew2001a,munro2001,roy2010}. Hence it would also be interesting to examine how coherence and Bell violation are connected to one 
another for the class of states considered in this paper. It has been observed that in most of the cases the computational techniques of measuring 
entanglement, coherence, mixedness et. al  involve computational basis. In this article, we shall make a comparative study of the quantification of 
these features with respect to both computational basis and Bell basis. Recently in an article \cite{kairon72021} it has been shown that for any general two 
qubit density matrix $\rho$, $\ell_{1}-$ norm of coherence, which is a measure of quantum coherence is always greater than or equal to concurrence of that density 
matrix. We will show by considering a class of mixed states that, the relation between coherence and concurrence is actually dependent on the choice of the basis under 
which they are measured and in a product basis where the multipartite basis is a product of the individual bases, then the $\ell_{1}-$ norm of coherence is greater than or equal to that 
of the entanglement as measured by concurrence.  When  we use an entangled basis like Bell basis this inequality does not hold.  This fact motivates us to explore how 
teleportation fidelities of the class of mixed states behave in accordance with coherence, entanglement and mixedness along with the Bell violation for these class of mixed states.

The paper is organized as follows. In section $2$ we discuss the quantifiers of coherence, entanglement and mixedness. It is to be pointed out that, although relative 
entropy of coherence is defined in the next section, in most of the calculations we have considered the $l_{1}-$ norm of coherence. The relative entropy and $\ell_{1}-$ norm, 
however, satisfies all the characteristics of good coherence measure. The mixedness of the class of states will also be studied and their relationship with both coherence 
and concurrence is investigated. The basic objective of studying any quantum mechanical state (pure or mixed) is to check its utility in information processing. 
Also we describe the two different classes of mixed states which we use in our investigations.  In section $3$, we compute the different measures for the two classes of 
states introduced in the previous section in both the computational basis and in the Bell basis. The  teleportation fidelity of these two classes of states is computed in the 
section  $4$ and the Bell-CHSH violation for these states is investigated in section $5$.  A comprehensive analysis of two qubit reduced density matrices derived from tripartite 
states is given in section $6$.  For all these states, we find quantum coherence, concurrence, mixedness, teleportation fidelity and the Bell-CHSH inequality.  Finally we present our 
conclusions in section $7$.

%
%
%
%

\section{Quantumness measures and Quantum States}
\subsection{Measures of coherence, entanglement and mixedness}
\subsubsection{Quantum coherence:} 
Quantum coherence originates  from the principle of superposition and hence it manifests itself in the off-diagonal elements of the density matrix corresponding 
to the state.  A  widely used quantum coherence quantifier is the $\ell_{1}$-norm measure and in our work we use the $C_{\ell_{1}}$ to represent this.   Since the 
quantum coherence is a basis dependent quantity, we fix the reference basis  $\ket{i}$ for a given quantum state.    The $l_{1}-$ norm of coherence is  defined 
as
\begin{eqnarray}
\label{l1norm}
C_{\ell_{1}}(\rho) = \sum_{i,j; i\ne j} \vert \rho_{ij}\vert, 
\end{eqnarray}
where $\rho_{ij} = \langle i\vert \rho\vert j\rangle$ is the matrix element corresponding to the $i^{th}$ row and $j^{th}$ column.  Hence if we consider a bipartite 
two qubit system in the standard computational basis, the reference basis is fixed at $\lbrace \vert 00\rangle, \vert 01\rangle, \vert 10\rangle, \vert 11\rangle\rbrace$.
An alternative method of quantifying coherence is the \textit{relative entropy of coherence}, denoted by $C_{r}$ and which is based on the relative entropy. 
Thus for a state $\rho$, the relative entropy of coherence is defined as 
\begin{eqnarray}
\label{relativeentropy1}
C_{r}(\rho) = \min_{\sigma \in \mathcal{I}}S(\rho \|\sigma),
\end{eqnarray}
where $S$ is the von-Neumann entropy. The minimum is taken over the set of the incoherent states $\mathcal{I}$ which are states without quantum coherence.  In Ref.\cite{baumgratz2014} it was proved that the 
expression for the relative entropy also reduces to the form
\begin{eqnarray}
\label{relativeentropy2}
C_{r}(\rho)= S(\rho_{d})-S(\rho),
\end{eqnarray}
where $\rho_{d}$ is the dephased state in the reference basis $\lbrace \vert i\rangle\rbrace$ i.e. the state obtained from $\rho$ by deleting all off-diagonal entries. Any  good measure of coherence must satisfy the properties for quantum coherence proposed in \cite{baumgratz2014} and both the 
$l_{1}-$ norm of coherence and relative entropy of coherence are the most general coherence monotones satisfying these properties.  One can construct a coherence measure 
using any distance measure if it satisfies the properties of positivity, monotonicity and convexity.  

An alternative framework for quantifying coherence has been proposed based on a natural property of coherence, the additivity of coherence for subspace-independent states. 
This is described by an operation independent equality rather than operation-dependent inequalities and therefore applicable to various physical contexts.  
This framework is also compatible with all the known results on coherence measures\cite{dm2016,dm2018}.   Recently a deterministic coherence distillation scheme has been proposed 
and the maximum number of maximally coherent states has been derived\cite{liu2019}.

\subsubsection{Concurrence:} 
Entanglement is an important quantum resource which arises due to nonlocal correlations between quantum sytems.  There are several measures for quantifying entanglement, 
like concurrence, relative entropy of entanglement, negativity and log negativity.  Of these measures concurrence is the widely accepted measure for two qubit systems and 
it works equally for pure and mixed states.  The concurrence of a quantum state $\rho$ \cite{wootters1998} is 
\begin{eqnarray}
\label{concurrence}
C(\rho) = \max \lbrace 0, \sqrt{\lambda_{1}}-\sqrt{\lambda_{2}}-\sqrt{\lambda_{3}}-\sqrt{\lambda_{4}}\rbrace,  
\end{eqnarray}
where $\lambda_{1}\ge\lambda_{2}\ge\lambda_{3}\ge\lambda_{4}$ are the eigenvalues of the matrix $\rho \tilde{\rho}$.  The spin-flipped density matrix $\tilde{\rho}$ is
\begin{eqnarray}
\label{spin-flipped}
\tilde{\rho} = (\sigma_{y}\otimes \sigma_{y})\rho^{*}(\sigma_{y}\otimes \sigma_{y}),
\end{eqnarray}
where $\sigma_{y}$ is the Pauli spin matrix in the $y$-basis and $\tilde{\rho}$ is in the same basis as $\rho$, and  $\rho^{*}$ is the complex conjugate of the density matrix $\rho$.   

\subsubsection{Mixedness:}
A quantum state is more often studied as an isolated system.  This is not exactly true and quantum states are in contact with an external environment which influences these 
states externally.  The environment causes a degradation of the quantumness of these states and causes them to decohere and become classical.  The quantumness of a state is 
maximal when it is pure and once it interacts with the environment, it usually loses its quantumness and consequently it appears mixed.  A completely  classical state is the maximal mixed state. 
When the state changes from being pure to becoming mixed, there is an entropy introduced which is quantified using the linear entropy.  For an arbitrary $d$-dimensional quantum mixed 
state $\rho$, the mixedness is defined using the normalized linear entropy  $L(\rho)$ and is defined as\cite{peters2004}
\begin{eqnarray}
\label{linearentropy1}
L(\rho)=\frac{d}{d-1}(1-{\rm Tr}(\rho^{2})).
\end{eqnarray}
Here the quantity ${\rm Tr} (\rho^{2})$  describes the purity of the quantum system.  For a two-qubit system, the value of $L$ ranges from $0$ to $1$. The entropy $L(\rho)= 0$ for any pure state, and the maximum value $L(\rho) = 1$ is attained for the maximally mixed state $\frac{I}{4}$.

%

\subsection{Quantum states}\label{sec:class}
The Bell states are a set of maximally entangled two qubit pure states and they are as follows: 
\begin{equation}
	\label{bellstates1}
	\vert \phi^{\pm}\rangle = \frac{1}{\sqrt{2}}\lbrace \vert 00\rangle_{AB} \pm \vert 11\rangle_{AB}\rbrace ;   \qquad  \qquad \qquad
	\vert \varphi^{\pm}\rangle = \frac{1}{\sqrt{2}}\lbrace \vert 01\rangle_{AB} \pm \vert 10\rangle_{AB}\rbrace.
\end{equation}
Here $A$ and $B$ denote the two particles Alice and Bob holding one qubit each.  These four states are orthogonal to each other
and form a basis which is in general referred to as the Bell basis.  The elements of the computational basis corresponding to the 
two qubit system can be represented in terms of the Bell basis elements as 
\begin{eqnarray}
	\label{bellstates2}
	\vert 00\rangle = \frac{1}{\sqrt{2}}\lbrace \vert \phi^{+}\rangle_{AB} + \vert \phi^{-}\rangle_{AB}\rbrace.\nonumber\\
	\vert 01\rangle = \frac{1}{\sqrt{2}}\lbrace \vert \varphi^{+}\rangle_{AB} + \vert \varphi^{-}\rangle_{AB}\rbrace.\nonumber\\
	\vert 10\rangle = \frac{1}{\sqrt{2}}\lbrace \vert \varphi^{+}\rangle_{AB} - \vert \varphi^{-}\rangle_{AB}\rbrace.\nonumber\\
	\vert 11\rangle = \frac{1}{\sqrt{2}}\lbrace \vert \phi^{+}\rangle_{AB} - \vert \phi^{-}\rangle_{AB}\rbrace.\nonumber\\
\end{eqnarray}
While we note that these two different bases can equally describe the Hilbert space of the two qubit systems, the computational 
basis is inherently separable whereas the Bell basis encodes nonlocal correlations in the basis vectors.  Following 
ref.\cite{bennett11996,horodecki11996} we can construct a class of mixed states as given below: 
\begin{eqnarray}
\label{class1}
\rho^{1} = p_{1}\vert \phi^{\pm}\rangle_{AB}\langle \phi^{\pm}\vert + (1-p_{1})\vert 00\rangle_{AB}\langle 00\vert\nonumber\\
\rho^{2} = p_{1}\vert \phi^{\pm}\rangle_{AB}\langle \phi^{\pm}\vert + (1-p_{1})\vert 11\rangle_{AB}\langle 11\vert\nonumber\\
\rho^{3} = p_{1}\vert \varphi^{\pm}\rangle_{AB}\langle \varphi^{\pm}\vert + (1-p_{1})\vert 01\rangle_{AB}\langle 01\vert\nonumber\\
\rho^{4} = p_{1}\vert \varphi^{\pm}\rangle_{AB}\langle \varphi^{\pm}\vert + (1-p_{1})\vert 10\rangle_{AB}\langle 10\vert,
\end{eqnarray}
and 
\begin{eqnarray}
\label{class2}
\varrho^{1} =  p_{2}\vert \phi^{\pm}\rangle_{AB}\langle \phi^{\pm}\vert + (1-p_{2})\vert 01\rangle_{AB}\langle 01\vert\nonumber\\
\varrho^{2} =  p_{2}\vert \phi^{\pm}\rangle_{AB}\langle \phi^{\pm}\vert + (1-p_{2})\vert 10\rangle_{AB}\langle 10\vert\nonumber\\
\varrho^{3} =  p_{2}\vert \varphi^{\pm}\rangle_{AB}\langle \varphi^{\pm}\vert + (1-p_{2})\vert 00\rangle_{AB}\langle 00\vert\nonumber\\
\varrho^{4} =  p_{2}\vert \varphi^{\pm}\rangle_{AB}\langle \varphi^{\pm}\vert + (1-p_{2})\vert 11\rangle_{AB}\langle 11\vert,
\end{eqnarray}
where $0 \leq p_{1} \le 1$ and $0 \leq p_{2} \le 1$  are the respective mixing parameters.  In future discussions  in this manuscript, we refer to the 
states described in Eq.(\ref{class1}) as the \textit{class-$1$} mixed states.  Similarly, the states given in Eq.(\ref{class2})
are referred to as the \textit{class-$2$} mixed states. Each one of these equations, i.e. Eq.(\ref{class1}) and Eq.(\ref{class2})
contain $16$ different mixed states and hence in total we have thirty  two different mixed states.  A quantum state in 
$\mathcal{C}^{2}\otimes \mathcal{C}^{2}$ or $\mathcal{C}^{2}\otimes \mathcal{C}^{3}$ is separable if and only if it obeys the PPT 
criteria\cite{peres1996}.  Any violation of this criteria in the $\mathcal{C}^{2}\otimes \mathcal{C}^{2}$ or $\mathcal{C}^{2}\otimes \mathcal{C}^{3}$
implies that the state is entangled.  Using this criteria we can show that the classes shown in Eq.(\ref{class1})and (\ref{class2}) 
are mixed entangled states under the conditions $0 < p_{1} < 1$ and $0 < p_{2} < 1$.  When  $ p_{1}$ and $p_{2}$ are zero, they 
are separable pure states and when $p_{1}$ and $p_{2}$ are both one, they  are pure entangled states.  In the following sections 
we investigate the coherence, concurrence and mixedness of these two classes of states.  For the sake of simplicity we denote  the 
computational basis with the notation $B_{1}=\lbrace \vert 00\rangle, \vert 01\rangle, \vert 10\rangle, \vert 11\rangle\rbrace$
and the Bell basis as $B_{2}=\lbrace \vert \phi^{+}\rangle, \vert \varphi^{+}\rangle, \vert \varphi^{-}\rangle, \vert \phi^{-}\rangle\rbrace$. 

\begin{figure}[h]
\includegraphics[width=\columnwidth]{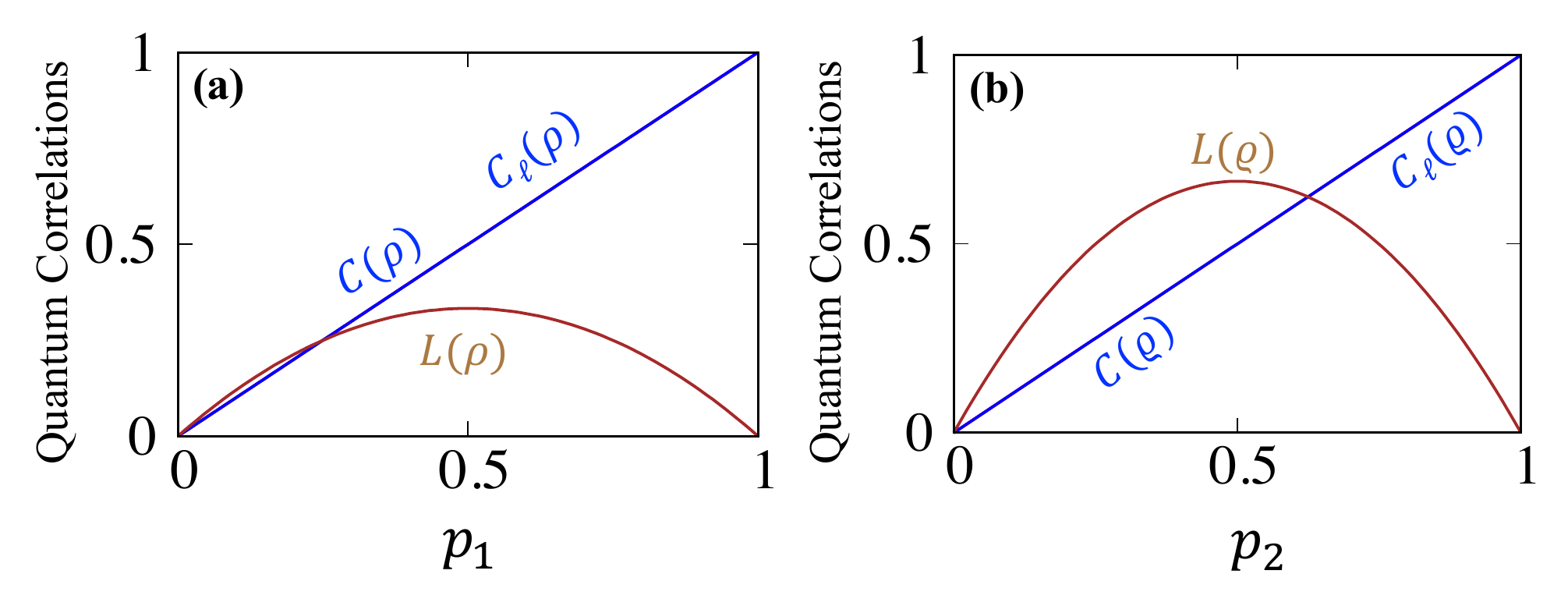}
\caption{The quantum coherence measured using the $\ell_{1}$ norm of coherence $C_{\ell_{1}}$, the entanglement measured using 
the concurrence and the mixedness estimated using the linear entropy $L$ are given for (a) \textit{class-$1$} type of states as a function 
of the probability $p_{1}$ and (b) \textit{class-$2$} type of states as a function of the probability $p_{2}$.  In this plot we use the 
computational basis for estimating the above mentioned quantifiers.   
 }
\label{fig1}
\end{figure}

%
%
%
%
\section{Choice of measurement basis}
\subsection{Measurement in the computational basis}
The quantum coherence, concurrence and the mixedness of the two classes of states {\it viz} \textit{class-$1$} states given through 
Eq.(\ref{class1}) and the  \textit{class-$2$} states given through Eq.(\ref{class2}) are computed in the computational basis.  
For the \textit{class-$1$} states the results obtained are 
\begin{eqnarray}
\label{class1result1}
C_{\ell_{1}}(\rho^{i}) &=& p_{1},\nonumber\\
C(\rho^{i}) &=& p_{1}, \nonumber\\
L(\rho^{i}) &=& \frac{4}{3}p_{1}(1-p_{1}), \forall i \in \{1,2,3,4\},
\end{eqnarray}
The results corresponding to the \textit{class-$2$} states are 
\begin{eqnarray}
\label{class2result1}
C_{\ell_{1}}(\varrho^{i}) &=& p_{2},\nonumber\\
C(\varrho^{i}) &=& p_{2},  \nonumber\\
L(\varrho^{i}) &=& \frac{8}{3}p_{2}(1-p_{2}), \forall i \in \{1,2,3,4\}.
\end{eqnarray}
From the analytic expressions we find that the coherence and concurrence are equal to each other.  This happens due to the following 
reason:  These two classes of states belong to the wider classification of states called the $X$-states in which the density matrix 
contains elements only along the main diagonal and the anti-diagonal.   A generic $X$-state has the following form as given below:
\begin{eqnarray}
\label{Xstate}
\rho^{m} = \left(%
\begin{array}{cccc}
	a & 0 & 0 & g\\
    0 &  b & f & 0\\
    0 &  f^{*} & c & 0\\
    g^{*} &  0 & 0 & d\\
\end{array}%
\right). 
\end{eqnarray}
For this state one can find the analytic expression of concurrence to be: 
\begin{equation}
C(\rho^{m}) = 2 ~{\rm max} \left\{ 0, |f|-\sqrt{ad}, |g|-\sqrt{bc} \right\}.
\label{concurrencexstate}
\end{equation}
For the \textit{class-$1$} set of states in Eq.(\ref{class1})  either the set $\{a, g, g^{*},d\}$ or the set $\{b, c,f, f^{*}\}$ which will 
give us the result to be either $C(\rho) = 2g$ or $C(\rho) = 2f$. The \textit{class-$2$} set of states in Eq.(\ref{class2}) has five elements
which again gives us the result of either $C(\rho) = 2g$ or $C(\rho) = 2f$ depending on which element is present.  The variation of 
quantum coherence, concurrence and mixedness is shown in Fig. (1) as a function of the mixing parameters.   For the \textit{class-$1$} 
set of states the results are shown in Fig. 1(a).  In Fig. 1(b), the results corresponding to \textit{class-$2$} set of states are given.  From 
both these plots we find that the coherence and the concurrence increase linearly with $p_{1}$ and $p_{2}$ as expected from the analytic
result.  The mixedness is maximum at $p_{1} = 0.5$ and $p_{2} = 0.5$ when the states are equally  present.  In the extreme cases when
$p_{1} = p_{2} = 0$ and $p_{1} = p_{2} = 1$ the states are pure and consequently the mixedness vanishes.

\subsection{Measurement in the Bell basis}
In this part we calculate the concurrence, quantum coherence and mixedness of the two classes of states namely the \textit{class-$1$} and the 
\textit{class-$2$} in the Bell basis.  It is well known that the Bell states are the set of maximally entangled states in the two qubit space.  They 
are linearly independent, hence forms a basis and hence any two-qubit state whether entangled or not can be expressed in the Bell basis.  In 
this subsection we look at the implications of computing the quantifiers like concurrence, coherence and mixedness in the Bell basis.  We begin
by expressing the density matrix in the Bell basis $\rho^{(b)}$.  To compute the concurrence we find $\tilde{\rho}^{(b)}$ as well in the same Bell 
basis using Eq.(\ref{spin-flipped}).  From the eigenvalues of the density matrix we can find the concurrence of the system.  For computing the quantum coherence, we use 
the density matrix in the Bell basis $\rho^{(b)}$ and find the distance to the closest diagonal state $\rho^{(b)}_{d}$ using the $\ell_{1}$-norm
coherence measure.  The mixedness of the state $\rho^{(b)}$ is found using the expression in Eq.(\ref{linearentropy1}).  	Below we summarize 
the results of our calculations.  For the \textit{class-$1$} type mixed states we have 
\begin{eqnarray}
\label{class1result2}
C_{\ell_{1}}(\rho^{i}) &=& 1- p_{1},\nonumber\\
C(\rho^{i}) &=& p_{1}, \nonumber\\
L(\rho^{i}) &=& \frac{4}{3}p_{1}(1-p_{1}), \forall i \in \{1,2,3,4\},
\end{eqnarray}
and for the \textit{class-$2$} type mixed states we have
\begin{eqnarray}
\label{class2result2}
C_{\ell_{1}}(\varrho^{i}) &=& 1- p_{2},\nonumber\\
C(\varrho^{i}) &=& p_{2}, \nonumber\\
L(\varrho^{i}) &=& \frac{8}{3}p_{2}(1-p_{2}), \forall i \in \{1,2,3,4\}.
\end{eqnarray}
\begin{figure}[h]
\includegraphics[width=\columnwidth]{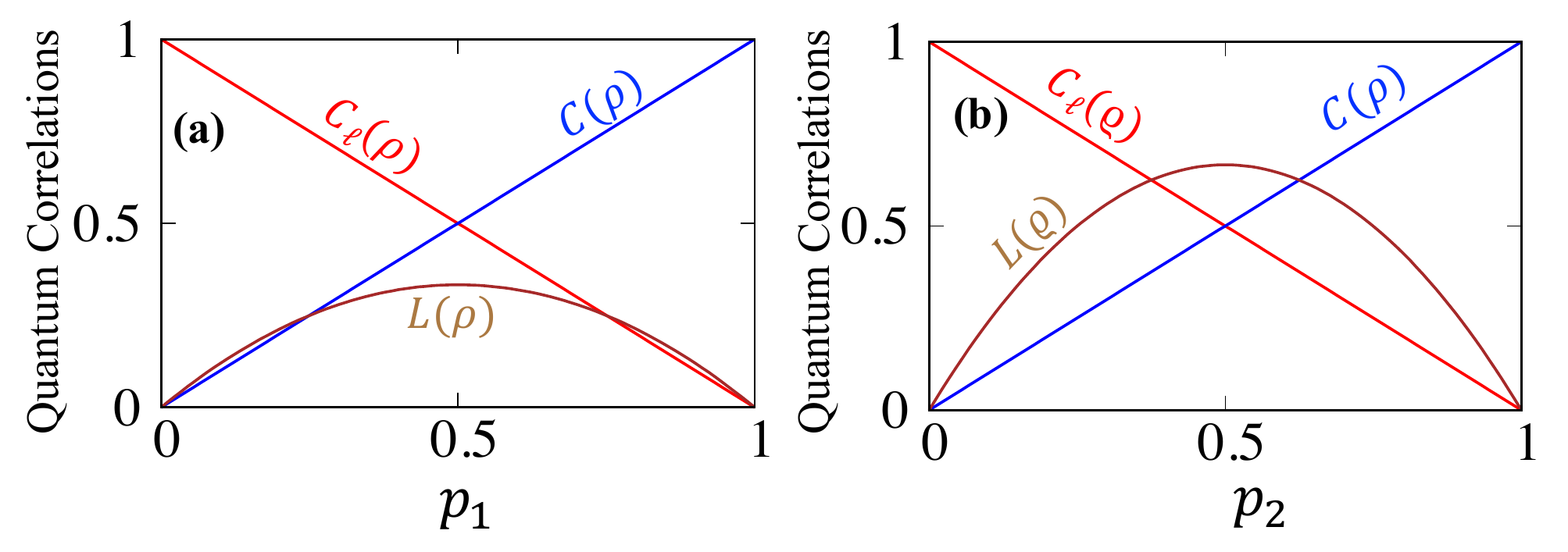}
\caption{The quantum coherence measured using the $\ell_{1}$ norm of coherence $C_{\ell_{1}}$, the entanglement measured using 
the concurrence and the mixedness estimated using the linear entropy $L$ are given for (a) \textit{class-$1$} type of states as a function 
of the probability $p_{1}$ and (b) \textit{class-$2$} type of states as a function of the probability $p_{2}$.  In this plot we use the 
Bell basis for estimating the above mentioned quantifiers.   }
\label{fig2}
\end{figure}
The variation of the quantum coherence, concurrence and mixedness with the mixing probabilities for the \textit{class-$1$} and \textit{class-$2$}
types of states are shown through the plots in Fig. 2(a) and Fig. 2(b) respectively. From the plots we can see that the coherence and the concurrence
are inversely proportional to each other.  When $p_{1}(p_{2})$ is zero, the coherence $C_{\ell_{1}}(\rho)$ ($C_{\ell_{1}}(\varrho)$) is maximum and 
goes to zero when $p_{1} (p_{2})$ is unity.  In the case of concurrence we find that $C(\rho)(C(\varrho))$ is minimum when the probabilities are zero and maximum when 
the probabilities are set to one.  This behavior is in contrast to the results obtained when these quantifiers are measured in the computational basis
where we observed that the coherence and concurrence varied in the same manner.  Now we explain the reason behind the contrasting results when 
we use these two different measurement bases.  The computational basis is a separable basis in the sense that the two qubit basis states can be 
written as a tensor product of the single qubit basis states.  Hence these basis states do not contain any type of quantum correlations or coherence within 
them.  Meanwhile if we look at the Bell basis the two qubit basis states are the entangled states and they cannot be written as tensor product of 
the individual basis states.  Hence these basis states inherently contain entanglement and coherence within them.  When we use the concurrence 
measure Eq.(\ref{concurrence}) to calculate the entanglement, we find the eigenvalues of the product of the density matrix and its spin-flipped 
version, with both these matrices expressed in the same Bell basis.  This computation gives us a basis independent result and hence the concurrence
is the same when we use either the computational basis or the Bell basis.  In fact the concurrence measure is an invariant quantity with respect to
the choice of the basis.  From the results on quantum coherence we find that the two different calculations arising from the computational basis 
and the Bell basis give opposite results.  This is because when we use the Bell basis for calculating quantum coherence, the density matrix is 
transformed to the Bell basis and then we use the same procedure as described through Eq.(\ref{l1norm}).  It is expected that this coherence 
need not be quantitatively same as the coherence measured using the computational basis, since quantum coherence as measured using 
Eq.(\ref{l1norm}) is a basis dependent quantity.  But what we are observing is a qualitative difference in behaviour where the relation 
between entanglement and coherence changes from being directly proportional to inversely proportional. This difference is because in the Bell 
basis there is coherence hidden within the basis which is not accounted for.  Hence in the calculation of coherence we should always choose 
a basis in which the multiqubit basis states are tensor product of the single qubit basis states to ensure no coherence is left unaccounted for.

The mixedness of these quantum states can be computed from the linear entropy of the system which is defined in Eq.(\ref{linearentropy1}). 
In general the mixedness can take values from $0$ to $1$.  From the plots of the two class of states we find that the maximal value of mixedness 
of the \textit{class-$1$} states is lower than that of the  \textit{class-$2$} states.  This is because the \textit{class-$1$} states have only two 
diagonal elements whereas the \textit{class-$2$} states have three diagonal elements and mixedness is the distance to the maximally mixed 
state $I/4$ where $I$ is the Identity matrix which has four diagonal elements.  When the mixing parameter is $0.5$, the \textit{class-$1$} states
and the \textit{class-$2$} states attain their maximal mixedness which are $L(\rho) = 1/3$ and $L(\varrho)=2/3$ respectively.  The qualitative and quantitative 
nature of mixedness is the same when it is computed either in the computational basis or in the Bell basis.

%
%
%
%
%
\section{Teleportation fidelity of the class of mixed states:}
Teleportation provides us a different notion of quantum inseparability.  In this method the information in a given quantum state is separated into the 
classical information and the quantum information and they are shared via a classical and quantum channel respectively. 
Later these quantum and classical parts can be combined to reconstruct the state with perfect fidelity.  For this process, we use an entangled 
pair of states as a quantum channel.  In general a singlet state $| \psi^{-} \rangle = \frac{1}{\sqrt{2}} (|01 \rangle - |10 \rangle)$ is used and the state is 
shared between Alice (sender) and Bob (receiver).  Now Alice decides to send an unknown reference state of the form $\alpha |0 \rangle + \beta |1 \rangle$ 
to Bob. To achieve this Alice performs Bell measurements jointly on her qubit and the reference qubit. Her measurement results are then conveyed to Bob by classical communication.   Finally Bob applies some unitary operators on his qubit depending upon Alice's measurement results.   In this process the unknown quantum state is destroyed at Alice's end and is recreated at Bob's end.  

The teleportation protocol as introduced by Bennett \textit{et. al} \cite{bennett21993} was based on using pure states as quantum channel.  
Later Popescu in Ref.\cite{popescu1994} found that the pairs of mixed states can also be used for teleportation, a feature known as probabilistic 
teleportation \cite{pankaj2002} or imperfect teleportation.  However there is a limitation to kind of quantum state which can be used 
in teleportation.  This limitation is captured by the measure called teleportation fidelity.  A quantum channel is useful for teleportation only if its fidelity 
exceeds $2/3$ which is the maximum possible fidelity achievable by means of Local Operations and Classical Communication (LOCC) and is known 
as the classical fidelity \cite{popescu1994,massar1995}.  A well known two qubit mixed state is the Werner state and has the form 
\begin{equation}
  \rho_{w} = \frac{1-F_{w}}{3} I_{4} + \frac{4 F_{w} - 1}{3} |\psi^{-} \rangle \langle \psi^{-} |
\end{equation}
where $|\psi^{-} \rangle = (|01 \rangle - |10 \rangle)/\sqrt{2}$ is the maximally entangled singlet and $F_{w}$ is the maximal singlet fraction in the
Werner state.  When the maximal singlet fraction $F_{w}$ lies in the range $\frac{1}{2} \leq F_{w} \leq \frac{3+\sqrt{2}}{4\sqrt{2}}$ the Werner state
satisfies the Bell-CHSH inequality, but the state is still entangled in this region.  This is because a quantum state which violates Bell-CHSH inequality
is said to exhibit nonlocality  a form of quantum correlation which is much stronger than entanglement and in the hierarchy of quantum correlations 
is rarer than entanglement \cite{radhakrishnan2019b}.  Hence Werner states can be used as quantum teleportation channel with average optimal 
fidelity exceeding $2/3$ even though it does not violate the Bell-CHSH inequality in the given region \cite{roy2010}.    

The utility of the class of maximally and non-maximally entangled states in quantum teleportation has been studied in Ref.\cite{roy2010}.  Below we 
examine the teleportation fidelity of the class of states defined in the previous section.  For an arbitrary density matrix $\rho$, the optimal teleportation 
fidelity is 
\begin{equation}
\label{teleportationfidelity}
f_{T}(\rho) = \frac{1}{2}\Bigg[1 + \frac{N(\rho)}{3}\Bigg],
\end{equation} 
where $N(\rho) = \sum_{i=1}^{3}\sqrt{u_{i}}$. Here, $u_{i}$'s are the eigenvalues of the matrix $T^{\dagger}T$. The elements of $T$,  denoted by $t_{st}$ are
\begin{equation}
\label{teleportationfidelity1}
t_{st} = Tr(\rho\sigma_{s}\otimes \sigma_{t}),
\end{equation}
and $\sigma_{j}$'s denote the Pauli spin matrices. In terms of the quantity $N(\rho)$, one can say that any mixed state $\rho$ is useful for standard teleportation
if and only if \cite{horodecki21996}  $N(\rho) > 1$.  For the \textit{class-$1$} states defined in Eq. (\ref{class1}), $\forall i$,  $N(\rho^{i}) > 1$ with $0 < p_{1}\leq 1$ and the 
optimal teleportation fidelity defined in Eq.(\ref{teleportationfidelity}) for these states is 
\begin{equation}
\label{telepfidclass1}
f_{T}(\rho^{i}) = \frac{2 + p_{1}}{3},
\end{equation}
Now for the \textit{class-$2$} states defined in Eq.(\ref{class2}), we have $\forall j$, $N(\varrho^{j}) > 1$  when $0.5 \leq p_{2} \leq 1$.  
The optimal teleportation fidelity of these states is 
\begin{equation}
\label{telepfidclass2}
f_{T}(\varrho^{j}) = \frac{1 + 2p_{2}}{3}. 
\end{equation}
These results hold both in the computational basis and Bell basis. It is easy to observe analytically that the teleportation fidelity  of the  \textit{class-$1$}  states always exceeds the classical 
teleportation value of $2/3$.  Hence these states can be used as teleportation channels for all values of mixing parameter. As for \text{class-$2$} states it is observed that in the range $0 \leq p_{2} < 1/2$ the teleportation 
fidelity of the states is less than the classical teleportation fidelity and is not suitable for quantum teleportation.  For the range $1/2 \leq p_{2} \leq 1$ however, 
the teleportation fidelity of the states is greater than the classical teleportation fidelity of $2/3$ and hence it is suitable for quantum teleportation.  
Hence we find the  \textit{class-$1$} states are better candidates for quantum teleportation in comparison to  \textit{class-$2$}.

%
%
%
%
%
\section{Bell-CHSH violation for the class of mixed states}
\label{BellCHSH}
The paradox which arose due to the Einstein-Podolsky-Rosen thought experiment also known as EPR paradox, where they concluded that quantum mechanical description of nature is incomplete since it goes against our common sense perceptions of locality and reality\cite{einstein1935}.  To verify the existence of non-local correlations, Bell proposed an inequality and a simpler form of this inequality is the 
CHSH inequality\cite{bell1964,clauser1969}.  The verification of the Bell-CHSH inequality for the \textit{class-$1$} and \textit{class-$2$} mixed states is defined in 
Eq.(\ref{class1}) and in Eq.(\ref{class2}) is carried out in this section.   

Any quantum state described by the density operator $\rho$ violates the Bell-CHSH inequality if and only if the following condition is satisfied: 
\begin{equation}
\label{bellchsh}
M(\rho) = \max_{i>j}(u_{i}+u_{j})>1.
\end{equation}
Here $u_{i}$'s are eigenvalues of the matrix $T^{\dagger}T$ \cite{horodecki31996}.  The elements of the matrix $T$ are defined in the 
Eq.(\ref{teleportationfidelity1}) and  $\sigma_{i}$'s are the Pauli matrices. 

\noindent \textit{Bell violation of the class-$1$ states in the computational basis:}  For the  \textit{class-$1$} states defined in Eq.(\ref{class1}), the 
eigenvalues of the matrix $T^{\dagger}T$ are $u_{1}=p_{1}^{2}+1$, $u_{2}=p_{1}^{2}+1$ and $u_{3}=2p_{1}^{2}$.   The state $M(\rho^{i})>1$ 
for $0<p_{1} \leq 1$, $\forall i$.  Hence we can conclude that the \textit{class-$1$} type of mixed states violates the Bell-CHSH inequality for 
all values of the mixing parameter.  

\noindent \textit{Bell violation of the class-$2$ states in the computational basis:}  The eigenvalues of the $T^{\dagger}T$ matrix are 
$v_{1}=p_{2}^{2}$, $v_{2}=p_{2}^{2}$ and $v_{3}=(2p_{2}-1)^{2}$.  Using this we compute the three possible different summations 
namely $e_{1}=v_{2}+v_{1} = 2p_{2}^{2}$, $e_{2}=v_{3}+v_{1}=5p_{2}^{2}-4p_{2}+1$ and $e_{3}=v_{3}+v_{2} = 5p_{2}^{2}-4p_{2}+1$. 
In the range $\frac{1}{3}<p_{2}<1$, $M(\varrho^{j}) = 2p_{2}^{2}$ and for the region $0<p_{2}<\frac{1}{3}$, $M(\varrho^{j})=5p_{2}^{2}-4p_{2}+1$
based on the maximum value.     We find that the quantity $M(\varrho^{j})=5p_{2}^{2}-4p_{2}+1 > 1$ for $0.8<p_{2}<1$.  Similarly, the quantity 
$M(\varrho^{j})=2p_{2}^{2}>1$ when $0.7071<p_{2}<1$. Hence we see that the class $2$ mixed states $\varrho^{j}$ violate Bell-CHSH inequality 
$0.7071<p_{2}<1$, $\forall j$.  While these quantum states are correlated for any value of $p_{2}$, the states have non-locality only in the 
region $0.7071<p_{2}<1$.  

From these observations we find that the \textit{class-$1$} type mixed states violate the Bell-CHSH inequality for all values of the mixing parameter 
$p_{1}$ .  In the case of the \textit{class-$2$} type mixed states, the state violates Bell inequality when $p_{2}>0.7071$, while for the region  $0.5<p_{2}\leq 0.7071$
the states satisfy the Bell-CHSH inequality and in both these cases these type of mixed states have a teleportation fidelity which is higher than the 
classical teleportation fidelity of $2/3$.   If we use the Bell basis the results obtained are consistent with those of the computational basis which is shown in 
this section.

%
%
%
%
%
\section{Mixed bipartite states derived from tripartite states}
The bipartite mixed states in Eq.(\ref{class1}) and Eq.(\ref{class2}) are probabilistic mixtures of two qubit pure states.  To achieve this 
experimentally one might have to generate pure states and it is well known that perfect generation of pure states is an experimental 
challenge.  An alternative is to generate tripartite quantum states and obtain the two qubit mixed states by tracing out one of the qubits. 
We use the bipartite mixed states obtained from the reduced states of some well known pure as well as three qubit mixed states.  
In the first subsection we look at the reduced two qubit density matrices obtained from pure three qubit systems. The second subsection 
presents results where the two qubit mixed states are obtained from three qubit mixed states.

%
\subsection{Bipartite mixed states derived from pure tripartite states}
In this part we consider three qubit pure entangled states and consider the reduced density matrices arising out of them by tracing 
out any one of the qubits.   The three qubit pure states can be entangled in two different ways namely the three way entangled $GHZ$
state\cite{greenberger1,greenberger2} and the two way entangled $W$ state\cite{w1}.  In the $GHZ$ state all the three qubits are entangled in such a way that loss of any one 
of the qubits destroys all the entanglement in the system.  When a qubit is traced out in a three qubit $W$ state the remaining 
bipartite system still has residual entanglement in the system.  These two states are classified into two distinct SLOCC 
(Stochastic Local Operations and Classical Communication) classes since a state in a given class cannot be transformed to a state 
in another class using only SLOCC operations\cite{Durr2000,Verstraete2002,Akimasa2004,Miyake2004}.  

\vspace{0.2cm}

\noindent{\textit{SLOCC class of states:}}  The GHZ state is tripartite maximally  entangled state in which the loss of even a single 
qubit makes the quantum state disentangled.  The GHZ state has the form $|GHZ \rangle  =  (|000 \rangle + |111 \rangle )/\sqrt{2}$
and for this state the two qubit reduced density matrix is 
\begin{equation}
\label{ghzreduced}
\rho^{g}_{AB} = \rho^{g}_{BC}  = \rho^{g}_{AC}  = \left(%
\begin{array}{cccc}
	\frac{1}{2} & 0 & 0 & 0\\
    0 &  0 & 0 & 0\\
    0 &  0 & 0 & 0\\
    0 &  0 & 0 & \frac{1}{2}\\
\end{array}%
\right),
\end{equation}
where $\rho^{g}_{AB} = {\rm Tr}_{C} \;  \rho^{g}_{ABC}$ ($\rho^{g}_{BC} = {\rm Tr}_{A} \;  \rho^{g}_{ABC}$, $\rho^{g}_{AC} = {\rm Tr}_{B} \;  \rho^{g}_{ABC}$) is the reduced two qubit density matrix and $\rho^{g}_{ABC} =  |GHZ \rangle \langle GHZ|$. 
Computing the various quantum correlations for this state (\ref{ghzreduced}) we find the concurrence $C =0$,  coherence $C_{\ell_{1}} = 0$ and mixedness $L = 2/3$.  
The teleportation fidelity of the quantum state $f_{T}(\rho_{AB}^{g}) = f_{T}(\rho_{BC}^{g}) = f_{T}(\rho_{CA}^{g}) = 0$.
Since there is no entanglement in the two qubit reduced density matrix of a $GHZ$ state, the state (\ref{ghzreduced}) is not useful for any kind of teleportation.

A  $W$ state is a three qubit entangled state, in which entanglement is shared only between the pairs of qubits and there is no genuine 
entanglement as in a $GHZ$ state.  The three qubit $W$ state reads: $|W \rangle  =  (|001 \rangle + |010 \rangle + |100 \rangle )/\sqrt{3}$.
The reduced density of the $W$ state is 
\begin{equation}
\label{wreduced}
\rho^{W}_{AB} = \rho^{W}_{BC}  = \rho^{W}_{AC}  = \left(%
\begin{array}{cccc}
	\frac{1}{3} & 0 & 0 & 0\\
    0 &  \frac{1}{3} & \frac{1}{3} & 0\\
    0 &  \frac{1}{3} & \frac{1}{3} & 0\\
    0 &  0 & 0 & 0 \\
\end{array}%
\right), 
\end{equation}
where $\rho^{W}_{AB} = {\rm Tr}_{C} \;  \rho^{W}_{ABC}$ ($\rho^{W}_{BC} = {\rm Tr}_{A} \;  \rho^{W}_{ABC}$, $\rho^{W}_{AC} = {\rm Tr}_{B} \;  \rho^{W}_{ABC}$) is the reduced two qubit density matrix and $\rho^{W}_{ABC} = |W \rangle \langle W|$
Representing the two qubit reduced density matrix $\rho^{w} = \rho^{W}_{AB} = \rho^{W}_{BC}  = \rho^{W}_{AC}$, we give the 
coherence, concurrence, mixedness and teleportation fidelity below: 
\begin{eqnarray}
\label{w2qubitreduced}
C_{\ell_{1}}(\rho^{w}) &=& \frac{2}{3},\nonumber\\
C(\rho^{w}) &=& \frac{2}{3}, \\
L(\rho^{w}) &=& \frac{16}{27}, \nonumber\\
f_{T}(\rho^{w}) &=& 7/9.
\end{eqnarray}
The bipartite reduced density matrices obtained from the $W$-states have a teleportation fidelity higher than $2/3$ and so they can be 
used in quantum teleportation. 

\vspace{0.2cm}

\noindent\textit{{$W \tilde{W}$ state:}} The three qubit $W \tilde{W}$ state is a pure state which is a linear superposition of the 
$W$ state and the $\tilde{W}$ state (which in fact is a spin flipped version of the $W$ state)\cite{radhakrishnan22020}.  The expression for this tripartite state reads: 
\begin{eqnarray}
\label{wwbar1}
\vert W \tilde{W}\rangle = \frac{1}{\sqrt{2}}(\vert W\rangle + \vert \tilde{W}\rangle).
\end{eqnarray}
where
\begin{eqnarray}
\label{wwbar2}
\vert W\rangle = \frac{1}{\sqrt{3}}(\vert 001\rangle + \vert 010\rangle + \vert 100\rangle), \nonumber\\
\vert \tilde{W}\rangle = \frac{1}{\sqrt{3}}(\vert 110\rangle + \vert 101\rangle + \vert 011\rangle).
\end{eqnarray}
The $W\tilde{W}$ state has its coherence distributed in both three way and two way manner, apart from this single qubit 
coherences are also present in the system.  For this system, we find the tangle $\tau = 1/3$ and hence we know that the 
three qubit entanglement is also present in the system.  The two qubit reduced density matrix also has entanglement suggesting
that the bipartite entanglement is also present in the system.  In the computational basis the two qubit reduced density matrix of this state is 
\begin{eqnarray}
\label{wwbarden1}
\rho^{W \tilde{W}}_{AB} = \rho^{W \tilde{W}}_{BC} = \rho^{W \tilde{W}}_{AC}=
\left(%
\begin{array}{cccc}
	\frac{1}{6} & 	\frac{1}{6} & 	\frac{1}{6} & 0\\ 
    	\frac{1}{6} &  	\frac{1}{3} & 	\frac{1}{3} & \frac{1}{6}\\
    	\frac{1}{6} &  	\frac{1}{3} & 	\frac{1}{3} & 	\frac{1}{6}\\
    0 &  	\frac{1}{6} & 	\frac{1}{6} & \frac{1}{6}\\
\end{array}%
\right),
\end{eqnarray}
This two qubit mixed state has coherence value of $C_{\ell_{1}} = 2$, concurrence $C = 1/3$,  mixedness $L = 10/27$ and
teleportation fidelity $f_{T} = 7/9$. To examine the Bell-CHSH inequality violation we compute $M(\rho^{W \bar{W}})$ (where 
$\rho^{W \tilde{W}} = \rho^{W \tilde{W}}_{AB} = \rho^{W \tilde{W}}_{BC} = \rho^{W \tilde{W}}_{AC}$) which comes out to be $8/9$.
As $M(\rho^{W \tilde{W}} )< 1$, the state satisfies Bell-CHSH inequality. 

\vspace{0.2cm}

\noindent\textit{{Star state:}}  The three qubit quantum states considered above {\it viz} $GHZ$, $W$ and the $W \tilde{W}$ states 
are all symmetric states in that the reduced density matrices are all identical irrespective of which qubit is traced over.  As an example 
of asymmetric states we consider the three qubit star states\cite{radhakrishnan22020}.  In these states there is a central qubit which is entangled to two 
peripheral qubits.  These peripheral qubits are not entangled between themselves and the form of a three qubit star state is:
\begin{eqnarray}
\label{star1}
\vert S\rangle_{ABC} = \frac{1}{2}[\vert 000\rangle + \vert 100\rangle + \vert 101\rangle + \vert 111\rangle].
\end{eqnarray}
When we trace out the central qubit, the reduced density matrix of the two qubit system is 
\begin{eqnarray}
\label{starAB}
\rho_{AB}^{s} = \left(%
\begin{array}{cccc}
	\frac{1}{4} & 	0 & 	\frac{1}{4} & 0\\
    	0 &  	0 & 	0 & 0\\
    	\frac{1}{4} &  	0 & 	\frac{1}{2} & 	\frac{1}{4}\\
    0 &  	0 & \frac{1}{4} & \frac{1}{4}\\
\end{array}%
\right).
\end{eqnarray}
Although the state has zero concurrence, it has finite amount of coherence $C_{\ell_{1}} = 1$. This bipartite partition is separable and hence is not 
of use in computing quantum teleportation fidelity.  If we trace out any  one of the peripheral qubit we get the following reduced density 
matrix: 
\begin{eqnarray}
\label{starBC}
\rho_{BC}^{s} = \rho_{AC}^{s} = \left(%
\begin{array}{cccc}
	\frac{1}{2} & 	\frac{1}{4} & 	0 & \frac{1}{4}\\
    	\frac{1}{4} &  	\frac{1}{4} & 	0 & \frac{1}{4}\\
    	0 &  	0 & 	0 & 	0\\
    \frac{1}{4} &  	\frac{1}{4} & 0 & \frac{1}{4}\\
\end{array}%
\right).
\end{eqnarray}
The state has a concurrence $C = 1/2$, coherence $C_{\ell_{1}} = 3/2$ and mixedness $L = 1/3$.  The teleportation fidelity of this state
is $(7 + 2 \sqrt{2} )/2$ which is higher than the classical teleportation fidelity of $2/3$ and hence can be used for quantum teleportation. 
These reduced density matrices satisfy the Bell-CHSH inequality.

%
\subsection{Bipartite mixed states derived from mixed tripartite states}
In  this subsection, we consider  bipartite mixed states obtained by tracing out one qubit in a three qubit mixed state which is a convex
combination of three qubit pure states.  We investigate various quantum features like quantum coherence, entanglement, mixedness and teleportation 
fidelity for these states.  
 
\begin{figure}[h]
\includegraphics[width=\columnwidth]{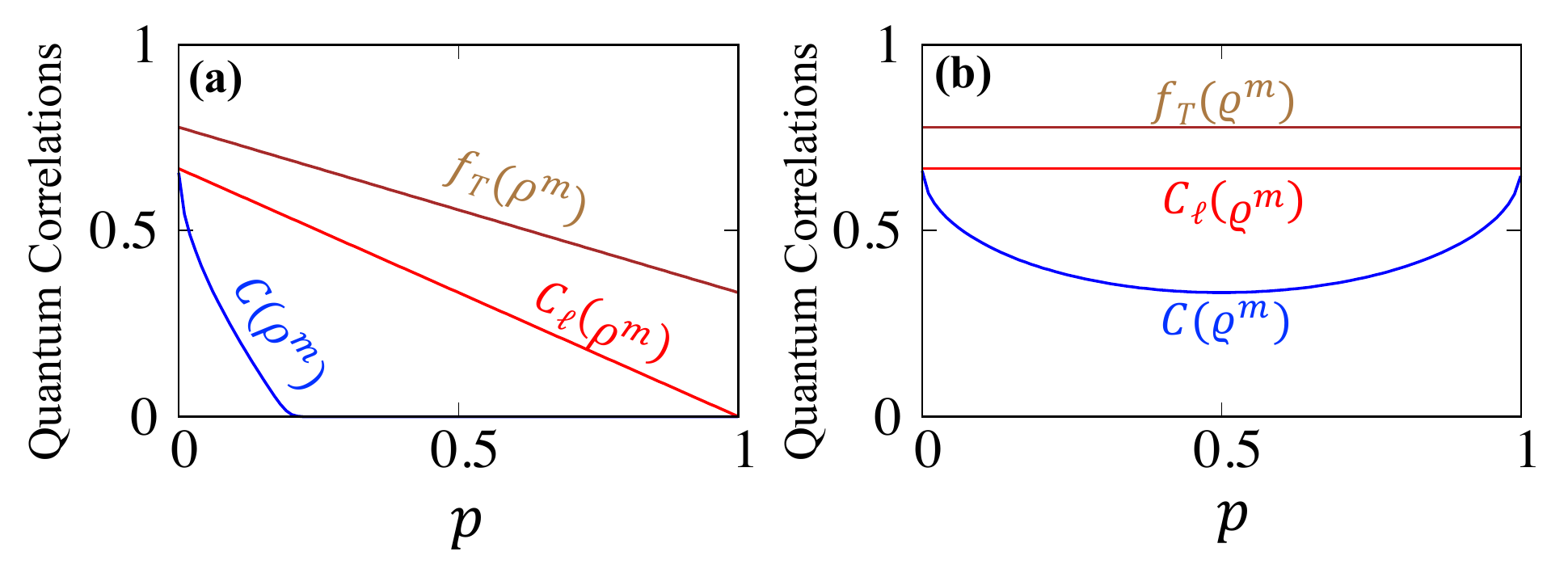}
\caption{The quantum coherence measured using the $\ell_{1}$ norm of coherence $C_{\ell_{1}}$, the entanglement measured using 
the concurrence and the teleportation fidelity $f_{T}$ are given for (a) mixture of $GHZ$ and $W$ states and (b) a mixture of $W$ state 
and $\tilde{W}$  states both as a function of the probability $p$.  }
\label{fig3}
\end{figure}

\noindent{\textit{Mixture of $GHZ$ and $W$ states:}} We consider the genuinely entangled three qubit $GHZ$ state and the bipartite 
entangled $W$ state and consider a tripartite state which is a convex mixture of both these states.  The form of this tripartite mixed 
state is
\begin{equation}
\label{mixghzwstate}
\rho^{gw}_{ABC} = p\vert GHZ\rangle_{ABC}\langle GHZ\vert + (1-p) \vert W\rangle_{ABC}\langle W\vert, ~~~~ 0\leq p \leq 1,
\end{equation}
where $p$ is the mixing parameter.  Now if we trace out any one of the qubits ($A$, $B$ or $C$) from the tripartite state Eq. (\ref{mixghzwstate}), 
the reduced density matrix $\rho^{gw}$ in the computational basis is
\begin{eqnarray}
\label{mixghzw2}
\rho^{gw} = \left(%
\begin{array}{cccc}
	\frac{p+2}{6} & 0 & 0 & 0\\
    0 &  \frac{1-p}{3} & \frac{1-p}{3} & 0\\
    0 &  \frac{1-p}{3} & \frac{1-p}{3} & 0\\
    0 &  0 & 0 & \frac{p}{2}\\
\end{array}%
\right). 
\end{eqnarray}
The coherence, concurrence and mixedness along with teleportation fidelity of the state (\ref{mixghzw2}) in computational basis are 
\begin{eqnarray}
\label{mixghzw3}
C_{\ell_{1}}(\rho^{gw}) &=& \frac{2(1-p)}{3},\nonumber\\
C(\rho^{gw}) &=& \frac{2(1-p)}{3}-2\sqrt{\frac{p(p+2)}{6}}, ~~ 0\leq p< 0.292\nonumber\\
L(\rho^{gw}) &=& \frac{2}{27}  \left( 8-13p^{2}+14p \right),\nonumber\\
f_{T}(\rho^{gw}) &=& \frac{7-4p}{9}, 0\leq p < \frac{1}{4}.
\end{eqnarray}
The application of this state as a teleportation channel has been investigated in \cite{roy2010} where it has been shown that the 
state is useful for quantum teleportation when $p$ lies between $0$ and $0.25$.  It is also to be noted that for $0.25 < p < 0.292$
though the state is entangled yet it cannot be used as a teleportation channel. The results of these investigations are shown in the 
plots in Fig. (\ref{fig3}). 

\vspace{0.2cm}

\noindent{\textit{Mixture of $W$ and $\tilde{W}$ states:}} Next we consider a convex mixture of $|W \rangle$ and $|\tilde{W} \rangle$ states
and investigate the teleportation capability  of two qubit reduced states obtained from these states. The expression for the three qubit 
mixed states read: 
\begin{eqnarray}
\label{mixwwbar1}
\varrho^{W \tilde{W}}_{ABC} &=& p \rho_{W} + (1-p) \rho_{\tilde{W}}\nonumber\\
&=& p \vert W\rangle_{ABC}\langle W\vert + (1-p) \vert \tilde{W}\rangle_{ABC}\langle \tilde{W}\vert, 
~~~~ 0\leq p \leq 1.
\end{eqnarray}
Tracing out any  one  of the three qubits i.e., $A$ or $B$ or $C$, the three qubit state $\varrho^{W \tilde{W}}_{ABC}$ reduces to a 
two qubit density matrix $\varrho^{w\tilde{w}}$ which is 
\begin{eqnarray}
\label{mixwwbar2}
\varrho^{w\tilde{w}} = \left(%
\begin{array}{cccc}
	\frac{p}{3} & 0 & 0 & 0\\
    0 &  \frac{1}{3} & \frac{1}{3} & 0\\
    0 &  \frac{1}{3} & \frac{1}{3} & 0\\
    0 &  0 & 0 & \frac{1-p}{3}\\
\end{array}%
\right),
\end{eqnarray}
where $\varrho^{w\tilde{w}} = \varrho^{W \tilde{W}}_{AB} = \varrho^{W \tilde{W}}_{BC}  = \varrho^{W \tilde{W}}_{AC}$. For this state, 
the coherence, concurrence, mixedness and teleportation fidelity are 
\begin{eqnarray}
\label{mixwwbar3}
C_{\ell_{1}}(\varrho^{w\tilde{w}}) &=& \frac{2}{3},\nonumber\\
C(\varrho^{w\tilde{w}}) &=& \frac{2}{3}-\frac{2}{3}\sqrt{p(1-p)},\nonumber\\
L(\varrho^{w\tilde{w}}) &=& \frac{8}{27} \left( 2-p^{2}+p  \right), \nonumber\\
f_{T}(\varrho^{w\tilde{w}}) &=& \frac{7}{9}.
\end{eqnarray}
Thus we find the teleportation fidelity to be greater than the classical teleportation fidelity of $2/3$ and hence the states 
can be useful in quantum teleportation.

\begin{table}[h!]
\begin{center}
\caption{Summary  of the quantum correlations, teleportation fidelity and Bell-CHSH inequality}
\label{table1}
\begin{tabular}{|c|c|c|c|c|}
\hline
\textbf{Quantum} & \textbf{Quantum} & \textbf{Concurrence} & \textbf{Mixedness} & \textbf{Teleportation}   \\
\textbf{states} & \textbf{Coherence $C_{\ell_{1}}$} &  $C$   & $L$  & \textbf{Fidelity $f_{T}$}   \\
\hline
\textit{class-$1$} states with &  $p_{1}$ & $p_{1}$ & $ 4 ( p_{1} (1-p_{1})) / 3$ & $(2+p_{1})/3$    \\
mixing parameter $p_{1}$ &   &   &   &    \\
\hline
\textit{class-$2$} states with &  $p_{2}$  & $p_{2}$ & $ 8 (p_{2} (1-p_{2})) /3$ & $(1 + 2 p_{2})/3$   \\
mixing parameter $p_{2}$ &   &   &   &    \\
\hline
$GHZ$ state  &  0  &  0  &  2/3 & 0 \\
2-qubit reduced form &   &   &   &  \\
\hline
$W$ state  &  2/3  &  2/3  &  16/27 & 7/9 \\
2-qubit reduced form &   &   &   &  \\
\hline 
$|W \bar{W} \rangle$ & $2$ & $1/3$ & $10/27$ & $7/9$ \\
2-qubit reduced form &   &   &   &  \\
\hline
Star state &  $2$ &  $3/2$ & $1/3$ & $(7 + 2 \sqrt{2})/2$ \\
2-qubit reduced form &   &   &   &  \\
\hline
$p \rho_{GHZ} +  (1-p) \rho_{W}$ &  $\frac{2}{3}(1-p)$ &  $ 2 \left( \frac{1-p}{3} - \sqrt{\frac{p(p+2)}{6}} \right)$ & $\frac{2}{27}( 8 - 13 p^{2} + 14 p )$ &  $(7 -4p)/9$  \\
2-qubit reduced form &   & $0 \leq p \leq 0.292$   &  &  $0 \leq p \leq 0.25$ \\
\hline
$p \rho_{W} + (1-p) \rho_{\tilde{W}}$ &   $2/3$ & $2/3 - 2 (\sqrt{p (1- p)})/3$ &  $\frac{8}{27} (2 - p^{2} + p)$ & $7/9$ \\
2-qubit reduced form &   &   &   &  \\
\hline
\end{tabular}
\end{center}
\end{table}

\section{Conclusion:} 
In this work, we study the quantum coherence and correlations of several bipartite mixed states.  Particularly, we find the quantum coherence, entanglement 
and mixedness of these quantum systems.  To estimate quantum coherence we use the well-known $\ell_{1}$ norm of coherence, in which we sum over
the off-diagonal elements of the two qubit density matrix.  The entanglement is measured using the concurrence measure which is suitable for both 
pure and mixed states.  Using the linear entropy we find the amount of mixedness in the states.  For the quantum states we consider a wider classes of 
states considering all possibilities.  Initially we consider the two qubit states created using a mixture of the maximally entangled Bell state with any one 
of the states from the computational basis.   Here we find that we can create two classes of states {\it viz} \textit{class-$1$} type and the  \textit{class-$2$} 
type of states.  In the \textit{class-$1$} type of states the basis vector being mixed with is also a constituent of the Bell state, whereas in the \textit{class-$2$} 
type of states it is not a constitutent of the Bell state with which it is being mixed.  Apart from this we also consider bipartite mixed states derived 
from either a tripartite pure or mixed state by tracing out one of their qubits.  A complete list of the states considered in this study are given in 
Table 1, where we have summarized the properties of each individual state. 

The quantum coherence, entanglement and mixedness of both the \textit{class-$1$} and \textit{class-$2$} type of states are computed both in the 
computational as well as in the Bell basis.  In the computational basis, the quantum coherence measured using the $\ell_{1}$-norm of coherence is equal 
to the entanglement estimated using the concurrence.  This is because we are using a specific class of $X$-states in which only two diagonal elements 
are present. The mixedness of the \textit{class-$1$} (\textit{class-$2$}) states varies  such that they are zero when either $p_{1}$ ($p_{2}$) is zero or 
when $p_{1}$ ($p_{2}$) is unity and the maximal value is attained midway when the probabilities are half.  When we estimate the coherence and concurrence
in the Bell basis we find them to have opposite behavior where the coherence is maximal when the concurrence is zero and the concurrence is maximal 
for vanishing coherence.  This is because in the Bell basis there is a coherence within the basis elements which is not accounted for when we compute the 
$\ell_{1}$-norm of coherence.  But concurrence measurements are basis independent and will give the same result irrespective of the basis chosen for 
the computation.  From this observation we learn that one should always use a incoherent basis (separable basis)  for the computation of quantum coherence in the 
system.  The use of either a coherent basis or an entangled basis might always lead to wrong results.  The mixedness in the bell basis is identical to the one 
obtained in the computational basis because the linear entropy is a basis independent evaluation.   Since the issue of the right kind of basis is settled, we 
computed the teleportation fidelity of both the \textit{class-$1$} and \textit{class-$2$} type of states and examined their suitability for quantum teleportation. 
The teleportation fidelity of the \textit{class-$1$} type of states is greater than the classical teleportation fidelity of $2/3$ for all values of the mixing parameter. 
Hence we conclude that these type of states can always be used for quantum teleportation.  In the case of the \textit{class-$2$} type of states, the teleportation 
fidelity is higher than the classical value only when $0.5 \leq p \leq 1$ and so these states can be used in quantum teleportation only in this range.  Next we verify
the Bell-CHSH inequality for both these class of states.  We find that the \textit{class-$1$} type of states violates the Bell-CHSH inequality for all values of 
the mixing parameter.  For the \textit{class-$2$} type of states we observe that the Bell-CHSH inequality is violated for $p > 0.7071$.  In the region 
$0.5 \leq p \leq 0.7071$, the states satisfy the Bell-CHSH inequality but they can still be used for quantum teleportation.  Finally we also investigate reduced
two qubit states obtained from the three qubit pure and mixed states. Here we consider a comprehensive set of pure states {\it viz} $GHZ$ state, $W$ state, 
$W \tilde{W}$ state, star state.  For the mixed states we consider the mixture of (i) $GHZ$ and $W$ state and (ii) $W$ and $\tilde{W}$ state.  The computed
correlations as well as their teleportation fidelity is listed in Table 1.  These results give us an idea of the suitability of various states for quantum teleportation. 
An interesting future work will be the examination of the teleportation fidelity of these different states under classical noise like telegraphic noise as well 
as dissipative and dephasing noise.  We are currently  working on these problems and the results will form the discussions of our future work.

\section*{Acknowledgements}
Chandrashekar Radhakrishnan was supported in part by a seed grant from IIT Madras to the Centre for Quantum Information, Communication and Computing.
Md. Manirul Ali was supported by the Centre for Quantum Science and Technology, Chennai Institute of Technology, India, vide funding 
number CIT/CQST/2021/RD-007.

\section*{Data Availability Statement}
The authors confirm that the data supporting the findings of this study are available within the article

\end{document}